\documentclass[10pt]{iopart}

\usepackage{iopams,setstack} 
\usepackage{cite} 
\usepackage{graphicx}
\usepackage{subcaption} 
\usepackage{bm}
\usepackage{amssymb}
\usepackage{hyperref} 
\usepackage{booktabs} 
\usepackage{braket}
\usepackage[margin=30mm]{geometry}
\usepackage{xcolor} 
\usepackage[british]{babel}
\hypersetup{colorlinks=true, anchorcolor=black} 

\newcommand{\preprint}{
  \setlength{\unitlength}{1mm}{\hbox{\begin{picture}(0,0)
        \put(100,10){\mbox{\footnotesize%
            ADP-26-18/T1315}}\end{picture}}}}

\begin{document}

\title[]{\preprint
Can multi-level Monte-Carlo address the sign problem in a finite-density Bose gas?}
\author{
Rajatha Ranpatige,
Finn M. Stokes
}
\address{ARC Special Research Centre for the Subatomic Structure of Matter (CSSM),
         Department of Physics, Adelaide University, SA, 5005, Australia.}
\ead{finn.stokes@adelaide.edu.au}

\begin{abstract}
	Numerical simulation of field theories with complex actions is frustrated by the sign problem, which
	causes an exponential growth in statistical sampling noise.
	A key example of such a theory is QCD at finite baryon chemical potential, which has important
	applications to the structure of atomic nuclei and neutron stars.
	Until now, the sign problem has precluded the possibility of first-principles numerical calculations
	that could provide key insight into these important physical systems.

	In this work, we propose an approach to avoiding this limitation by applying a reweighting procedure
	coupled with a multi-level Monte-Carlo sampling algorithm that offers an exponential enhancement of
	statistics to counteract the sign problem.
	We test the approach on a simpler model that displays the same sign problem at finite chemical potential.
	Our results demonstrate the promise of the proposed approach, significantly increasing the range over which precise calculations are possible, and providing improvements all the way up to the phase transition.
\end{abstract}

\noindent Keywords: lattice field theory, finite density, Monte-Carlo methods, sign problem

\section{Introduction}
Field theories with a complex action are extremely difficult to simulate, as the imaginary part
introduces a nontrivial complex phase into the exponential weight.
This phase prevents us from treating the weight as a probability for Monte-Carlo importance
sampling of the path integral.
This is known as the sign problem or complex action problem.
It has frustrated numerical efforts to understand the physics of large numbers of interacting
particles, such as in highly correlated condensed matter systems, atomic nuclei, neutron stars,
and quark-gluon plasma.

Much effort has been invested to search for solutions to the sign problem that would unlock
these important areas of study for over 40 years.
Many approaches have been proposed, for example
the complex Langevin approach~\cite{Seiler:2012wz,Sexty:2013ica,Aarts:2014bwa,Fodor:2015doa,
Sexty:2019vqx,Kogut:2019qmi,Scherzer:2020kiu, Ito:2020mys,Tsutsui:2025jez},
imaginary chemical potential~\cite{deForcrand:2002hgr,DElia:2002tig,DElia:2009pdy,DElia:2016jqh,
Cea:2014xva,Bonati:2014kpa,Cea:2015cya,Bonati:2015bha,Bellwied:2015rza,Bonati:2018nut,
Lombardo:2006yc,Guenther:2017hnx,Alba:2017mqu,Vovchenko:2017xad,Borsanyi:2018grb,Bellwied:2019pxh,
Borsanyi:2020fev,Cohen:2025ahp},
reweighting~\cite{Vladikas:1987tt,Barbour:1997ej,Fodor:2001au,Fodor:2001pe,Fodor:2004nz,
Giordano:2019gev,Borsanyi:2021hbk},
Taylor series expansion~\cite{Allton:2002zi,Allton:2005gk,Gavai:2003mf,Gavai:2004sd,Gavai:2008zr,
MILC:2008reg,Borsanyi:2011sw,Borsanyi:2012cr,Bellwied:2015lba,Giordano:2019slo,Borsanyi:2024xrx,
Ding:2015fca,Bazavov:2017dus,HotQCD:2018pds,Bazavov:2020bjn},
and Lefschetz thimble methods~\cite{Cristoforetti:2012su,Cristoforetti:2013wha,Alexandru:2015xva,
Alexandru:2016lsn,Nishimura:2017vav}.
However, it has been proven that this problem is NP-hard~\cite{Troyer:2004ge}, and so a general
solution is unlikely to exist.
In fact, none of those approaches have succeeded in solving quantum chromodynamics (QCD) at
high baryon chemical potential and low temperature, the exact conditions found within atomic
nuclei and neutron stars.
As a result, there have been no first-principles determinations of heavy nuclear structure or
neutron star physics directly from the physics of the strong force.
In this work we search for an approach to tackle this sign problem in this specific region
of particular interest, at low temperatures and high densities.

In QCD at zero temperature and non-zero chemical potential, it is expected that QCD undergoes
a transition to nuclear matter at a critical chemical potential \(\mu_c\simeq M_N / 3\), that
is one third of the nucleon mass.
Below this critical chemical potential, physical observables should not change with varying
\(\mu\), implying that the various effects of increasing \(\mu\) must exactly cancel out.
This seemingly-improbable cancellation is known as the ``Silver Blaze''
problem~\cite{Cohen:2003kd}.
It has been shown that this problem is intimately related to the sign problem in
QCD~\cite{Osborn:2004rf, Osborn:2005ss, Akemann:2004dr}.

In this work we demonstrate that a multi-level Monte-Carlo approach to reweighting may
offer a resolution to the sign problem at high densities and low temperatures.
This is demonstrated in a simplified scenario of a relativistic Bose gas that displays the same
``Silver Blaze'' phenomenon and sign problem.

\section{Relativistic Bose gas}
We consider a complex scalar field with a quartic interaction term in the presence of a chemical
potential \(\mu\).
Following the conventions of Ref.~\cite{Aarts:2009hn}, we write the continuum
Euclidean space Lagrangian density as
\begin{eqnarray}
	\mathcal{L}_\mathrm{cont.}(x) &= \partial_\nu \phi^*(x)\,\partial^\nu \phi(x) + \left(m^2 - \mu^2\right)\phi^*(x)\phi(x)\\
	& {}+ \mu \left(\phi^*(x) \partial_4 \phi(x) - \partial_4 \phi^*(x) \phi(x)\right) + \lambda {\left(\phi^*(x)\phi(x)\right)}^2\,.
\end{eqnarray}
We require \(m^2 > 0\) and \(\lambda\) sufficiently small that the \(\mathbb{Z}_2\) symmetry \(\phi\to -\phi\)
is unbroken for zero \(\mu\).
This symmetry remains unbroken for any sufficiently small value of \(\mu\).
The Lagrangian density is complex for any \(\mu > 0\), and obeys the relation \(\mathcal{L}(-\mu) = \mathcal{L}^*(\mu)\).

Discretising this theory on a \(d\)-dimensional space-time lattice, we obtain the discretised Lagrangian
\begin{equation*}\label{eq:action}
	\mathcal{L}(x) = (2d + m^2) \phi^*_x \phi_x - \sum_{\nu=1}^4 \left(\phi^*_xe^{-\mu\delta_{\nu,4}}\phi_{x+\hat{\nu}} + \phi^*_{x+\hat{\nu}}e^{\mu\delta_{\nu,4}}\phi_x\right) + \lambda {\left(\phi^*_x\phi_x\right)}^2\,.
\end{equation*}
We use a \(d=4\) lattice of size \(V = L_s\times L_s\times L_s\times L_\tau\),
where \(L_s\) is the length, or number of sites in each spatial direction and \(L_\tau\) is the length
in the temporal direction.

This complex Lagrangian poses a problem when we try to take the path integral using Monte-Carlo methods.
Consider the path integral
\begin{equation*}
	\braket{O} = \frac{1}{Z} \int \mathcal{D}\phi \, O[\phi] e^{-S[\phi]}\,,
\end{equation*}
where \(Z = \int \mathcal{D}\phi \, e^{-S[\phi]}\).
Since the action \(S = \sum_x \mathcal{L}(x)\) is complex, the \(e^{-S}\) factor is not purely real and thus can't be interpreted as a probability.
To resolve this, we start by splitting the resulting action into the real and imaginary parts
\begin{equation*}
	S = S_{\mathbb{R}} + i S_{\mathbb{I}}\,,
\end{equation*}
where
\begin{eqnarray}
	S_{\mathbb{R}} &= \sum_x\Big[(2d + m^2) \phi^*_x \phi_x - \sum_{i=1}^3 \left(\phi^*_x\phi_{x+\hat{i}} + \phi^*_{x+\hat{i}}\phi_x\right)\\
	&\qquad\qquad- \cosh(\mu) \left(\phi^*_x\phi_{x+\hat{4}} + \phi^*_{x+\hat{4}}\phi_x\right) + \lambda {\left(\phi^*_x\phi_x\right)}^2\Big]\,,\mathrm{\ and}\\
	S_{\mathbb{I}} &= -i \sum_x \sinh(\mu) \left(\phi^*_x\phi_{x+\hat{4}} - \phi^*_{x+\hat{4}}\phi_x\right) \,.
\end{eqnarray}
Both \(S_{\mathbb{R}}\) and \(S_{\mathbb{I}}\) are purely real.
Then the path integral becomes
\begin{equation*}
	\braket{\mathcal{O}} = \frac{1}{Z} \int \mathcal{D}\phi \, \mathcal{O}[\phi] e^{-S_\mathbb{R}[\phi]} e^{-iS_\mathbb{I}[\phi]}\,.
\end{equation*}
The \(e^{-S_\mathbb{R}}\) factor is now real, and can be treated as a probability for Monte-Carlo sampling.
However, we have a leftover complex phase \(e^{-iS_\mathbb{I}}\) that we need some way to deal with.

\begin{figure}
    \centering
    \includegraphics[width=0.8\linewidth]{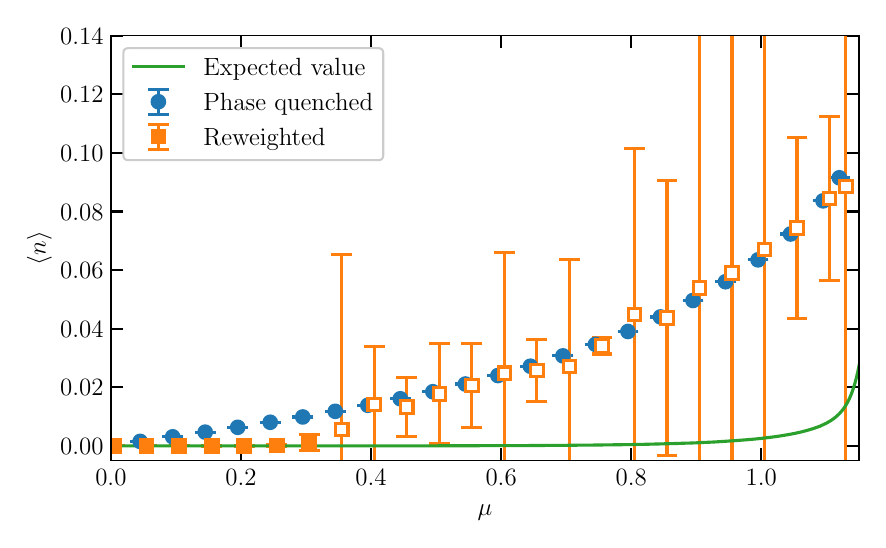}
    \caption{Plot of the density $\braket{n}$ as a function of the chemical potential $\mu$, comparing phase-quenched and reweighted determinations to the expected behaviour.
    These results come from 1,000,000 Monte-Carlo samples on an \(8\times8\times8\times8\) lattice.
    The phase-quenched results, represented by circular markers, sit significantly above
    the expected curve, from Ref~\cite{Aarts:2009hn}.
    The reweighted results, represented by square markers, agree precisely with the expected curve at low chemical potential, at or below $\mu=0.3$.
    Above, the statistical noise grows significantly, and the uncertainties become unreliable due to their distribution becoming non-Gaussian.
    This region is indicated by open markers.
    Points have been offset slightly in $\mu$ for readability.}
    \label{fig:phase-comparison}
\end{figure}

One approach is to neglect it and obtain what is known as the phase quenched expectation value
\begin{equation*}
	\braket{\mathcal{O}}_{PQ} = \frac{1}{Z_{PQ}} \int \mathcal{D}\phi \, \mathcal{O}[\phi] e^{-S_\mathbb{R}[\phi]}\,,
\end{equation*}
where \(Z_{PQ} = \int \mathcal{D}\phi\, e^{-S_\mathbb{R}[\phi]}\).
This approach is numerically straightforward, but introduces an uncontrolled systematic effect due to
the neglected phase.
This can be seen in Figure~\ref{fig:phase-comparison}, where we plot the expectation value of the density \(\braket{n} = (1/V) \partial \ln Z / \partial\mu\) as a function of the chemical potential. The phase-quenched curve sits well above the expected density value.

Another approach is to treat the phase as part of the observable~\cite{Vladikas:1987tt}, in an approach known as reweighting.
Then the expectation value becomes
\begin{equation*}
	\braket{\mathcal{O}} = \frac{\braket{\mathcal{O}e^{-iS_\mathbb{I}[\phi]}}_{PQ}}{\braket{e^{-iS_\mathbb{I}[\phi]}}_{PQ}}\,,
\end{equation*}
where the \(\braket{e^{-iS_\mathbb{I}[\phi]}}_{PQ}\) accounts for the difference between the phase-quenched and full partition functions.
With enough Monte-Carlo samples, this approach will give the correct result.
However, due to cancellations from the different phases of different \(\phi\) configurations, the
statistical uncertainties from the Monte-Carlo sampling blow up and the result becomes extremely noisy, as can be seen in Figure~\ref{fig:phase-comparison}.
Above a chemical potential of \(\mu=0.3\), the sign problem leads to extremely slow convergence of the ensemble average to the central limit theorem.
This results in non-Gaussian behaviour for the statistical uncertainty, which can be identified by long tails in the bootstrap resampled distribution.
The points in this region are plotted with open markers and their uncertainties should be treated as unreliable, as they have a large error-on-the-error, often more than 100\%.

\section{Multi-level Monte-Carlo\label{sec:multi-level}}
\subsection{Formulation\label{sec:multi-level:formulation}}
To resolve this statistical noise problem, we turn to multi-level algorithms~\cite{Luscher:2001up,Meyer:2002cd,
DellaMorte:2007zz,DellaMorte:2008jd,DellaMorte:2010yp}, which have been shown to provide an incredible
increase in statistical precision.
This approach relies on finding a factorisation of the action and the observable under consideration
into components that can be Monte-Carlo sampled independently.
This factorisation is straightforward for local theories like the scalar field theory we consider here.
For QCD, integrating out the fermion fields introduces non-local contributions that complicate this
factorisation, but a domain decomposition technique has been developed that resolves this issue~\cite{
Ce:2016idq,Ce:2016qto}.

To compute the expectation value of an operator \(\mathcal{O}\), given our action \(S\), we need to
compute the path integral
\begin{equation}
	\braket{\mathcal{O}} = \int \mathcal{D}\phi \, \mathcal{O}[\phi] \exp(-S[\phi])\,.
\end{equation}
Our multi-level approach begins by decomposing the lattice \(\Gamma\) into
\(D^{(1)}\) domains \(\Gamma_1\), \(\Gamma_2\), \ldots, and a remainder \(\Omega\) such that the action can be written as
the sum of terms \(S_i\) that only depend on \(\Gamma_i\) and \(\Omega\),
\begin{equation}
	S[\phi_\Gamma] = \sum_i S_i[\phi_{\Gamma_i}, \phi_\Omega]\,,
\end{equation}
and the operator can be written as the sum of products of similar terms
\begin{equation}
	\mathcal{O}[\phi_\Gamma] = \sum_j \prod_{i=1}^{D^{(1)}} \mathcal{O}_{ji}[\phi_{\Gamma_i}, \phi_\Omega]\,,
\end{equation}
where \(\phi_X\) are the fields restricted to the domain \(X\).
The operator is decomposed in an unusual way here, as a sum of products, in order
to accommodate a broad range of potential operators.
In most cases, the operator will simply be a product of terms on the different domains
(i.e. the sum only runs over a single \(j\)), or a sum of terms on the different domains
(i.e. \(\mathcal{O}_{ji}=1 \,\forall\, i \ne j\)).

For a local action and local operators, like the ones we consider here, it is sufficient for the \(\Gamma_i\) to be
disjoint domains separated by a boundary region \(\Omega\) that is at least one lattice site thick.

With such a factorisation in hand, we can rearrange the path integral as
\begin{equation}
	\braket{\mathcal{O}} = \int \mathcal{D}\phi_{\Omega} \sum_j \prod_{i=1}^{D^{(1)}} \int \mathcal{D}\phi_{\Gamma_i}
		\mathcal{O}_{ji}[\phi_{\Gamma_i}, \phi_\Omega] \exp(-S_i[\phi_{\Gamma_i}, \phi_\Omega])\,.
\end{equation}
We can evaluate this nested path integral through a multi-level Monte-Carlo integration approach.
Starting with a conventional Monte Carlo sampling which we call level 0, we obtain snapshots
\(\phi^{(0),k}_\Gamma\) of the full field, distributed according to the full action
\begin{equation}
	p^{(0)}[\phi_\Gamma] = \exp(-S[\phi_\Gamma]) = \prod_{i=1}^{D^{(1)}} \exp(-S_i[\phi_{\Gamma_i}, \phi_\Omega])\,.
\end{equation}
If we take \(N^{(0)}\) of these level-0 samples we can compute the conventional Monte-Carlo estimate for
the expectation value
\begin{equation}
	\braket{\mathcal{O}} \simeq \frac{1}{N^{(0)}} \sum_{k^{(0)}=1}^{N^{(0)}} \mathcal{O}[\phi^{(0),k^{(0)}}_\Gamma]
	= \frac{1}{N^{(0)}} \sum_{k^{(0)}=1}^{N^{(0)}} \sum_j \prod_{i=1}^{D^{(1)}} \mathcal{O}_{ji}[\phi^{(0),k^{(0)}}_{\Gamma_i}, \phi^{(0),k^{(0)}}_\Omega]\,.
\end{equation}
The statistical variance of such an estimate scales as
\begin{equation*}
	{\left(N^{(0)}\right)}^{-1}\,,
\end{equation*}
but a multi-level approach can do much better.

If we consider a single level-0 sample, \(\phi^{(0),k^{(0)}}_\Gamma\), we can perform a level-1 sampling
in the following way:
We ``freeze'' the field within the boundary region \(\phi^{(0),k^{(0)}}_\Omega\), and sample the individual \(\Gamma_i\)
domains independently, with a fixed boundary set by our frozen \(\phi_\Omega\).
The fields \(\phi^{(1),k^{(0)},k^{(1)}_i}_{\Gamma_i}\) on each domain are weighted by
\begin{equation}
	p^{(1)}[\phi^{(1),k^{(0)},k^{(1)}_i}_{\Gamma_i}] = \exp(-S_i[\phi^{(1),k^{(0)},k^{(1)}_i}_{\Gamma_i}, \phi^{(0),k^{(0)}}_\Omega])\,.
\end{equation}
Then, taking \(N^{(1)}\) of these level-1 samples for each domain and for each \(k^{(0)}\) gives a set of estimates
\begin{equation}
	\braket{\mathcal{O}}^{(1),k^{(0)}} = \sum_j \prod_{i=1}^{D^{(1)}} \frac{1}{N^{(1)}} \sum_{k^{(1)}_i=1}^{N^{(1)}} \mathcal{O}_{ji}[\phi^{(0),k^{(0)},k_i^{(1)}}_{\Gamma_i}, \phi^{(0),k^{(0)}}_\Omega]\,,
\end{equation}
which approximate the desired expectation value.
Because the domains are sampled independently, and the product of their contributions is taken,
we effectively consider all possible combinations of the individual samples, so the variance of the estimate has the potential
to scale as
\begin{equation*}
	{\left(N^{(1)}\right)}^{-D^{(1)}}\,,
\end{equation*}
a significant improvement over the level-1 estimate.
However, these individual estimates have a systematic bias from the frozen boundaries.
By also averaging over the level-0 samples we obtain the estimate
\begin{equation}
	\braket{\mathcal{O}} \simeq \frac{1}{N^{(0)}} \sum_{k^{(0)}=1}^{N^{(0)}} \sum_j \prod_{i=1}^{D^{(1)}} \frac{1}{N^{(1)}} \sum_{k^{(1)}_i=1}^{N^{(1)}} \mathcal{O}_{ji}[\phi^{(0),k^{(0)}}_{\Gamma_i}, \phi^{(0),k^{(0)}}_\Omega]\,,
\end{equation}
which is free from this bias.
Then, contributions that don't depend on \(\Omega\) have variances that potentially scale as
\begin{equation*}
	{\left(N^{(0)}\right)}^{-1} {\left(N^{(1)}\right)}^{-D^{(1)}}\,,
\end{equation*}
and the additional bias correction associated with \(\Omega\) scales as
\begin{equation*}
	{\left(N^{(0)}\right)}^{-1}\,,
\end{equation*}
while the cost of the calculation scales roughly as \(N^{(0)}N^{(1)}\).
As a result, it is possible to achieve a dramatic reduction in uncertainty for the same computational cost.

Optimising this approach then requires maximising the number of domains, while minimising the
\(\Omega\)-dependent contributions.
Often, improving one of these requires worsening the other, so there is often a trade-off to find the correct
balance.
It is possible to take the multi-level approach further, by dividing each \(\Gamma_i\) into \(D^{(2)}\) domains \(\Gamma_{ij}\),
and the remainder \(\Omega_i\), then performing a level-2 sampling on the individual \(\Gamma_{ij}\).
This can be continued indefinitely to level 3, 4, and so on.
This allows different trade-offs between the number of domains, and the boundary dependence at different levels,
with each successive level increasing the number of domains at the cost of more boundary effects.
This allows for the sources of variance to be balanced against one another to maximise the final precision
for a given computational cost.

\subsection{Decompositions\label{sec:decompositions}}
In this work we test a variety of approaches for decomposing the lattice into domains for level-1 multi-level sampling
and compare their effectiveness.
Different approaches can change the trade-off between number of domains and number of frozen sites, depending
on the specific geometry of the original lattice.
As we shall see in Section~\ref{sec:results}, the specific action and operators under consideration can favour
certain geometries over others, even when the number of domains and number of frozen sites are identical.

For an \(L_s\times L_s\times L_s \times L_t\) periodic lattice \(\Gamma\), we can construct a
class of decompositions by taking axis-aligned parallel slices through the lattice.

For example, we define \(T^\delta(D)\) to be the decomposition defined by freezing \(D\) equally-spaced slices
of thickness 1 at specific \(t\) values.
\(\delta\) is a four-vector offset that labels the different possible translations of the decomposition
scheme.
For this decomposition to fit neatly on the periodic lattice, we require that \(L_t\) is divisible by \(D\),
so that our slices fit neatly on the lattice with an equal spacing \(\Delta = L_t / D\)
between them.
This gives
\begin{equation*}
	\Omega(T^\delta(D)) = \{x \,|\, (x_4 - \delta_4) \;\mathrm{mod}\; \Delta = 0\}\,,
\end{equation*}
and the resulting domains are
\begin{equation*}
	\Gamma_i(T^\delta(D)) = \{x \,|\, i\Delta < (x_4 - \delta_4) <(i+1)\Delta\}\,.
\end{equation*}
Similar axis-aligned decompositions (\(X^\delta(D)\), \(Y^\delta(D)\), and \(Z^\delta(D)\))  can be defined by
slicing across one of the spatial dimensions instead.

We can generalise these decompositions to include diagonal slices that cut across multiple axes simultaneously,
which we label \(S_{v}^\delta(D)\).
For simplicity, we focus on slices orthogonal to some four-dimensional normal vector \(v\) with \(v_\mu \in {\{-1, 0, +1\}}\).
We define \(L_v\) to be the greatest common divisor of the axis lengths for which \(v_\mu\) is nonzero
(e.g. \(L_{[0,0,1,1]} = \gcd(L_s, L_t)\), and \(L_{[1,-1,1,0]} = L_s\)).
We require that \(L_v\) is divisible by \(D\), so our slices have equal spacing \(\Delta = L_v / D\).
This gives a frozen region of
\begin{equation*}
	\Omega(S_{v}^\delta(D)) = \{x \,|\, (x - \delta)\cdot v \;\mathrm{mod}\; \Delta = 0\}\,,
\end{equation*}
and the resulting domains are
\begin{equation*}
	\Gamma_i(S_{v}^\delta(D)) = \{x \,|\, i\Delta < (x - \delta)\cdot v <(i+1)\Delta\}\,.
\end{equation*}
The frozen sites \(\Omega\) take the form of \(D\) three-dimensional slices that total
to \(1 / \Delta\) of the total lattice volume.

\begin{figure}[tbp]
    \centering
    \begin{subfigure}[t]{0.3\textwidth}
        \centering
        \includegraphics[width=\textwidth]{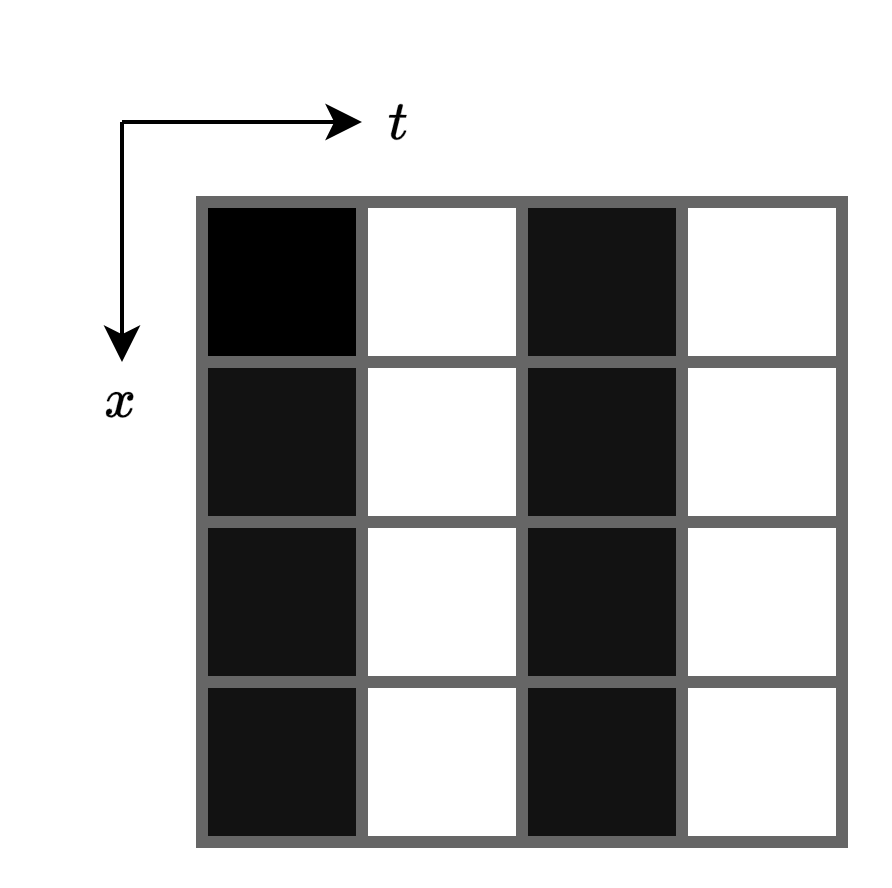}
        \caption{Slicing in T: $T^0(2)$}
        \label{fig:orthogonal1}
    \end{subfigure}
    \hfill
    \begin{subfigure}[t]{0.3\textwidth}
        \centering
        \includegraphics[width=\textwidth]{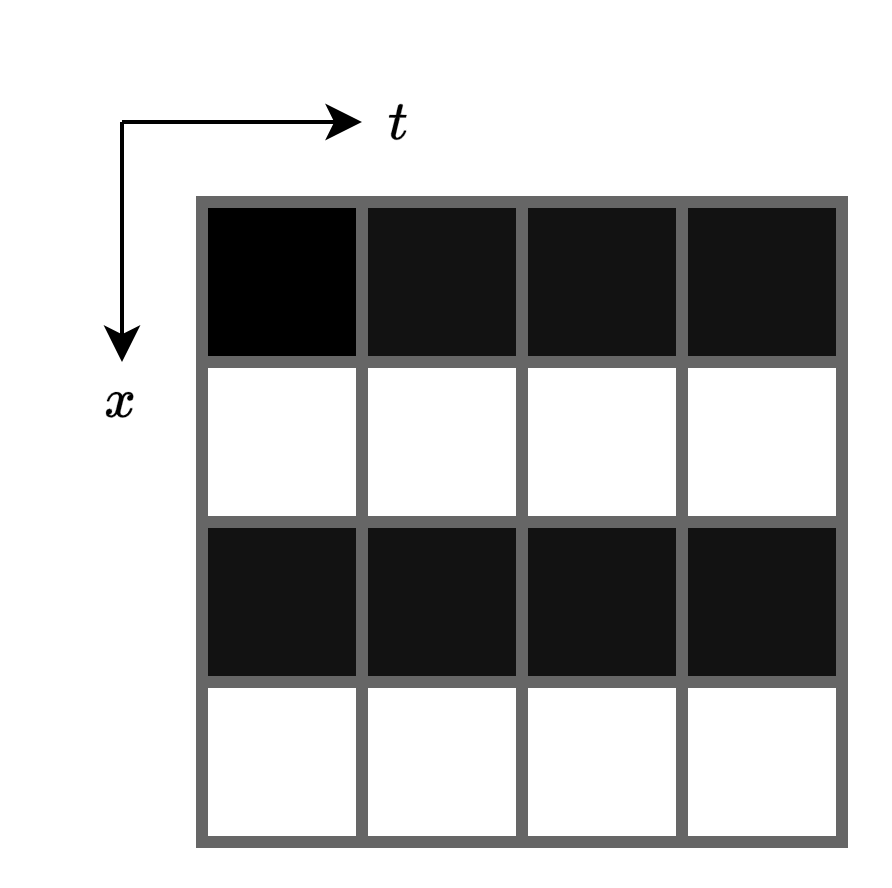}
        \caption{Slicing in X: $X^0(2)$}
        \label{fig:orthogonal2}
    \end{subfigure}
    \hfill
    \begin{subfigure}[t]{0.3\textwidth}
        \centering
        \includegraphics[width=\textwidth]{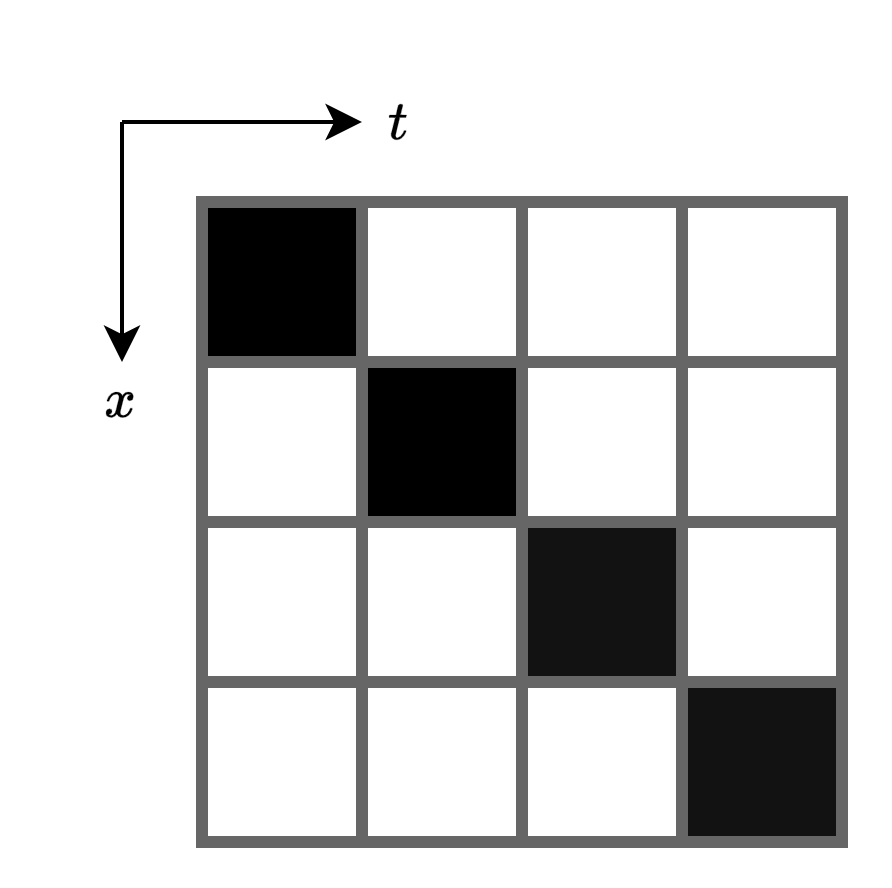}
        \caption{\centering Slicing in direction $(-1,0,0,1)$: $S^0_{[-1,0,0,1]}(1)$}
        \label{fig:diagonal}
    \end{subfigure}
    \caption{Some examples of different ways the lattice can be sliced into subdomains, illustrated on a two-dimensional slice through a \(4\times4\times4\times4\) lattice.
    Black squares indicate frozen sites, and white squares the non-frozen sites, forming two domains in (a) and (b) and one in (c).}
    \label{fig:both}
\end{figure}

Our earlier axis-aligned decompositions are simply \(X^\delta(D) = S_{\hat{x}}^\delta(D)\),
\(Y^\delta(D) = S_{\hat{y}}^\delta(D)\), \(Z^\delta(D) = S_{\hat{z}}^\delta(D)\), and
\(T^\delta(D) = S_{\hat{t}}^\delta(D)\).
Some examples of these decompositions are shown in Figure~\ref{fig:both}.

We can also combine multiple of these decompositions to subdivide the lattice further, which we denote with \(\otimes\).
The decomposition \(S_v^\delta(D)\otimes S_{v'}^\delta(D')\) freezes
\begin{equation*}
	\Omega(S_v^\delta(D)\otimes S_{v'}^\delta(D')) = \Omega(S_v^\delta(D)) \cup \Omega(S_{v'}^\delta(D'))\,.
\end{equation*}
We might expect that the domains are then the intersections of each pair of domains from the two original decompositions.
However, there are often more domains than that due to the way the slices intersect with one another when wrapping around the
periodic boundary of the lattice.

We can repeat this process to subdivide the lattice even further.
For example, \(X^0(2)\otimes Y^0(2)\otimes Z^0(2)\otimes T^0(2)\) subdivides the lattice into sixteen domains
with the same shape as the original lattice, but approximately half the size in every dimension. 

\begin{figure}[tb]
    \centering
    \includegraphics[width=0.3\linewidth]{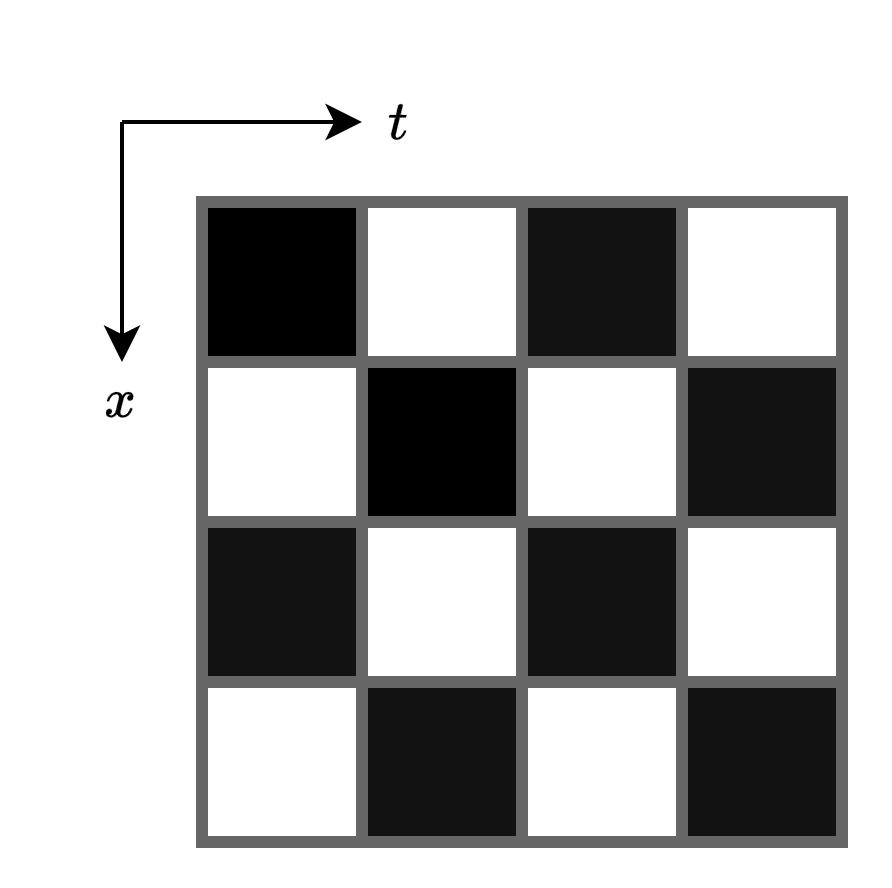}
    \caption{A checkerboard decomposition $C^0 = S^0_v(2)$, illustrated on a two-dimensional slice through a \(4\times4\times4\times4\) lattice.
    Black squares indicate frozen sites, and white squares the non-frozen sites, forming eight independent domains on each slice, for a total of 128 domains.}
    \label{fig:checkerboard}
\end{figure}

If we consider a lattice with even \(L_s\) and \(L_t\), and
take a fully diagonal slicing such as \(S^\delta_{[1,1,1,1]}(D)\) to the extreme, where
\begin{equation*}
	D = L_{[1,1,1,1]} / 2 = \gcd(L_s, L_t) / 2\,,
\end{equation*}
then half the sites will be frozen, forming an alternating
``checkerboard'' with the non-frozen sites, as shown in Figure~\ref{fig:checkerboard}.
Since every neighbour of every non-frozen site is in \(\Omega\), we can treat each of them as
an independent domain without freezing any additional sites.
This gives us a total of \(L_x\cdot L_y\cdot L_z\cdot L_t / 2\) independent domains with
half of the lattice sites frozen.
This is the maximum number of domains we can achieve for our action.
We call this decomposition a checkerboard decomposition and label it \(C^\delta\).

Table~\ref{tab:decompositions} shows the characteristics of an assortment of these decompositions
on three simple lattices: \(6\times6\times6\times6\), \(8\times8\times8\times8\), and \(16\times16\times16\times16\).
For two domains, the axis-aligned decompositions like \(T^\delta(2)\)
and the more diagonal decompositions like \(S_{[1,1,1,1]}^\delta(2)\) perform equally well.
However, for higher numbers of domains, it becomes optimal to combine multiple orthogonal slicings.
When combining slicings, more diagonal decompositions become advantageous due to the way
they wrap around the periodic boundaries and intersect with one another.
To achieve the largest number of domains the checkerboard is by far the best option.

\begin{table}[htbp]
	\begin{center}
\begin{tabular}{llll}
	Decompositions & Lattice Size & \# Domains & Frozen \\
	\hline
	\(S_v^\delta(2)\) & \(6\times6\times6\times6\) & 2 & 33\% \\
	\(S_v^\delta(2)\) & \(8\times8\times8\times8\) & 2 & 25\% \\
	\(S_v^\delta(2)\) & \(16\times16\times16\times16\) & 2 & 13\% \\
	\(S_v^\delta(4)\) & \(8\times8\times8\times8\) & 4 & 50\% \\
	\(S_v^\delta(4)\) & \(16\times16\times16\times16\) & 4 & 25\% \\
	\(S_v^\delta(8)\) & \(16\times16\times16\times16\) & 8 & 50\% \\
	\(X^\delta(2)\otimes Y^\delta(2)\) & \(8\times8\times8\times8\) & 4 & 44\% \\
	\(S_u^\delta(2)\otimes S_w^\delta(2)\) & \(8\times8\times8\times8\) & 8 & 38\% \\
	\(S_a^\delta(1)\otimes S_b^\delta(1)\otimes S_c^\delta(1)\otimes S_d^\delta(1)\) & \(6\times6\times6\times6\) & 16 & 40\% \\
	\(S_a^\delta(1)\otimes S_b^\delta(1)\otimes S_c^\delta(1)\otimes S_d^\delta(1)\) & \(8\times8\times8\times8\) & 16 & 34\% \\
	\(S_a^\delta(1)\otimes S_b^\delta(1)\otimes S_c^\delta(1)\otimes S_d^\delta(1)\) & \(16\times16\times16\times16\) & 16 & 21\% \\
	\(S_a^\delta(2)\otimes S_b^\delta(2)\otimes S_c^\delta(2)\otimes S_d^\delta(2)\) & \(8\times8\times8\times8\) & 256 & 44\% \\
	\(S_a^\delta(2)\otimes S_b^\delta(2)\otimes S_c^\delta(2)\otimes S_d^\delta(2)\) & \(16\times16\times16\times16\) & 256 & 34\% \\
	\(S_a^\delta(4)\otimes S_b^\delta(4)\otimes S_c^\delta(4)\otimes S_d^\delta(4)\) & \(16\times16\times16\times16\) & 4096 & 44\% \\
	\(C^\delta\) & \(6\times6\times6\times6\) & 648 & 50\% \\
	\(C^\delta\) & \(8\times8\times8\times8\) & 2048 & 50\% \\
	\(C^\delta\) & \(16\times16\times16\times16\) & 32768 & 50\%
\end{tabular}
	\end{center}
	\caption{\label{tab:decompositions}Comparison of different decomposition schemes for some example lattices,
	presenting the number of domains and percentage of frozen sites for each.
	Better precision is attained for larger numbers of domains and a smaller percentage of sites frozen.
	\(v\) is any normal vector with \(v_\mu \in {\{-1, 0, +1\}}\) and \(\delta\) is any offset.
	\(u = [1, 1, 0, 0]\) and \(w = [1, -1, 0, 0]\) are a pair of orthogonal vectors, and
	\(a = [1, 1, 1, 1]\), \(b = [1, 1, -1, -1]\), \(c = [1, -1, 1, -1]\) and \(d = [1, -1, -1, 1]\)
	are four orthogonal vectors. \(S\), \(X\) and \(Y\) are parallel slicings, and \(C\) is the
	checkerboard decomposition all described in Sec.~\ref{sec:decompositions}.}
\end{table}

\section{Results\label{sec:results}}
We performed simulations at \(m=\lambda=1\) on three different sizes of lattice: \(6\times6\times6\times6\), \(8\times8\times8\times8\), and \(16\times16\times16\times16\) and compared reweighted calculations of the density \(\braket{n}\) at chemical potentials from \(\mu=0\) up to the critical potential \(\mu_c=1.15\), using a variety of multi-level strategies.
To provide a fair comparison, we kept a constant number of Monte-Carlo samples \(N=N^{(0)} N^{(1)}\), while varying the number of level-0 and level-1 samples \(N^{(0)}\) and \(N^{(1)}\) for various different decomposition schemes.
This ensures that the computational cost would be the same or better than a standard Monte-Carlo sampling with \(N\) samples.
Since the different domains are completely independent at level 1, the calculation of the level-1 samples has significantly less communication required, so will likely be more efficient than level-0 samples.
We found that different schemes performed better for different lattice volumes, but the same general trends were consistent.

\begin{figure}[tbp]
    \centering
    \includegraphics[width=0.8\linewidth]{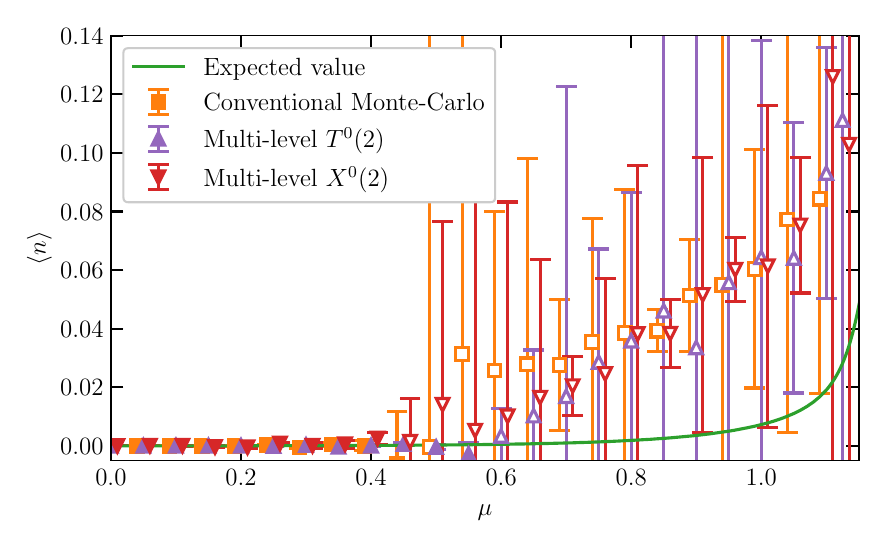}
    \caption{Comparison of different multi-level decompositions for a \(6\times6\times6\times6\) lattice.
    The plot shows densities obtained through the reweighting approach.
    Temporal slicing (\(T^0(2)\), upward triangles) maintains precise results (filled points) up to $\mu=0.55$, further than spatial slicing (\(X^0(2)\), downward triangles), which behaves similarly to conventional Monte-Carlo, reaching $\mu=0.4$ before non-Gaussianity sets in (open points).
    At large $\mu$, there is little difference between the two approaches.
    Points have been offset slightly in $\mu$ for readability.}
    \label{fig:L6666}
\end{figure}

On the smallest (\(6\times6\times6\times6\)) lattice, the potential benefits of multi-level were limited, and the optimal approach we found was the simplest decomposition into two regions.
Because the lattice volume is small, the surface-to-volume ratio is large, and so any slices freeze a notable fraction of the sites, so decompositions with a minimal number of slices are best.
If we compare two orthogonal decompositions: \(X^0(2)\) and \(T^0(2)\), with \(N^{(0)} = 200\) and \(N^{(1)} = 200\), we might expect them to be equivalent, but it turns out slicing in the time direction is more effective, as shown in Fig.~\ref{fig:L6666}.
If we inspect the forms of the phase factor and the density
\begin{eqnarray}
    e^{iS_{\mathbb{I}}} &= \exp\!\left(\sum_x\sinh(\mu) \left(\phi^*_x\phi_{x+\hat{4}} - \phi^*_{x+\hat{4}}\phi_x\right)\right) \,,\ \mathrm{and}\\
    n &= \sum_x \left[\sinh(\mu)\left(\phi_x^*\phi_{x+\hat{4}} + \phi_x\phi_{x+\hat{4}}^*\right) - \cosh(\mu)\left(\phi_x^*\phi_{x+\hat{4}} - \phi_x\phi_{x+\hat{4}}^*\right)\right]\,,
\end{eqnarray}
these combine neighbouring pairs in the time direction, but not in any spatial direction.
This means that when we slice in the time direction (\(T^\delta(D)\)), terms depending on a frozen site also depend on a non-frozen site.
In contrast, when we slice in a spatial direction (e.g. \(X^\delta(D)\)), there are half as many terms depending on a frozen site, but they all depend on two frozen sites, and have no non-frozen part.
Our testing consistently shows that the partially frozen terms in time-sliced decompositions lead to much more improvement from the multi-level sampling than the fewer doubly-frozen terms in space-sliced decompositions.
This also extends to diagonal slicings: \(S_v^\delta(D)\) has the same half-frozen behaviour as \(T^\delta(D)\) if and only if \(v_4 \ne 0\).
This means that appropriately chosen diagonal slicings have another advantage on top of those already discussed in Sec.~\ref{sec:decompositions}: they can be combined with orthogonal diagonal slices that all have this half-frozen property.
This is not the case for \(T^\delta(D)\), as all of the orthogonal slicings have the double-frozen property because they are necessarily spatial.

\begin{figure}[tb]
    \centering
    \includegraphics[width=0.8\linewidth]{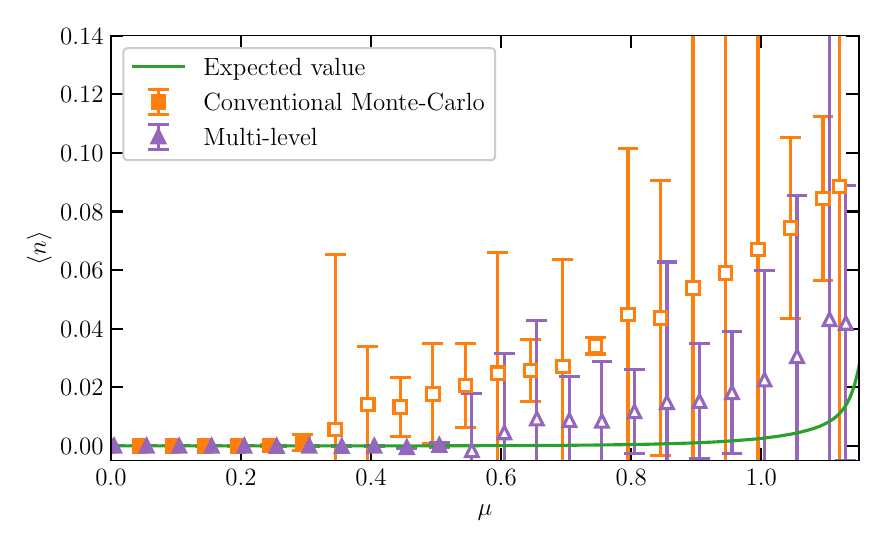}
    \caption{Comparison of multi-level (triangles) and conventional Monte-Carlo sampling (squares) for a \(8\times8\times8\times8\) lattice.
    The plot shows densities obtained through the reweighting approach.
    The multi-level sampling uses a decomposition based on intersecting diagonal slices in four orthogonal directions: \(S^0_{[1,1,1,1]}(2)\otimes S^0_{[1,1,-1,-1]}(2)\otimes S^0_{[1,-1,1,-1]}(2)\otimes S^0_{[1,-1,-1,1]}(2)\).
    It achieves precise results (solid points) up to significantly higher $\mu$ than the conventional Monte-Carlo ($\mu=0.5$ compared to $\mu=0.3$).
    In the region with non-Gaussian uncertainties (open points), the multi-level results have central values closer to the true values, and significantly smaller uncertainties than the conventional Monte-Carlo sampling.
    Points have been offset slightly in $\mu$ for readability.}
    \label{fig:L8888}
\end{figure}

On the middle (\(8\times8\times8\times8\)) lattice, there is enough volume available that more complicated decompositions become optimal.
The best approach we identified was a combination of four orthogonal diagonal slicing patterns: \(S^0_{[1,1,1,1]}(2)\otimes S^0_{[1,1,-1,-1]}(2)\otimes S^0_{[1,-1,1,-1]}(2)\otimes S^0_{[1,-1,-1,1]}(2)\), with \(N^{(0)} = 100,000\) and \(N^{(1)} = 10\).
As shown in Fig.~\ref{fig:L8888}, this approach significantly increases the range of \(\mu\) for which a precise result with well-controlled uncertainties is possible, from \(0 \le \mu \le 0.3\) to \(0 \le \mu \le 0.5\).
It also provides a noticeable improvement all the way up to the phase transition, although the statistical uncertainties are still non-Gaussian above \(\mu=0.5\).
An investigation of the phase contributions from the level-0 samples suggests that unlike the conventional Monte-Carlo sampling, this non-Gaussianity seems to come from poor overlap between the sampled, phase-quenched, distribution and the true underlying distribution, rather than cancellations between different contributions.
So here, the sign problem seems to have been replaced with an overlap problem, a known issue with reweighting techniques.

\begin{figure}[tbp]
    \centering
    \includegraphics[width=0.8\linewidth]{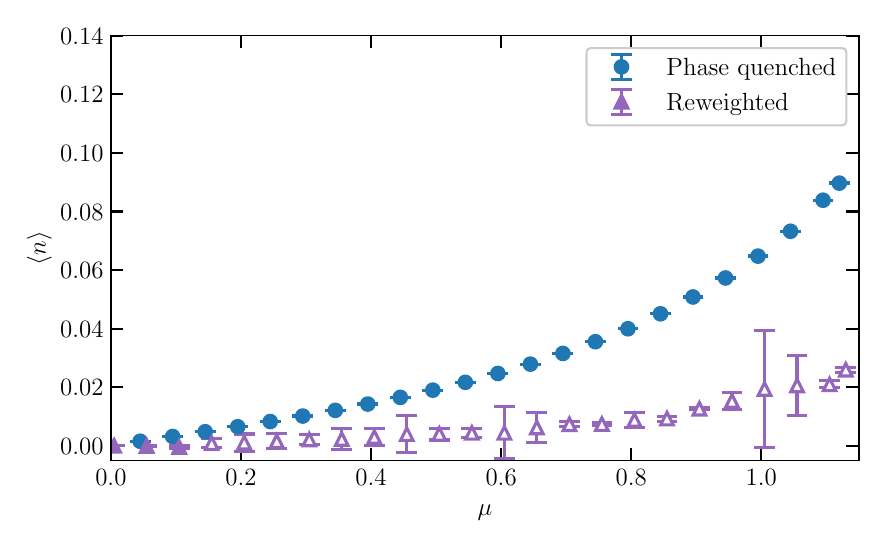}
    \caption{Multi-level reweighted density (triangles) compared to the phase quenched results (circles) for a \(16\times16\times16\times16\) lattice.
    The multi-level sampling uses a decomposition based on intersecting diagonal slices in four orthogonal directions: \(S^0_{[1,1,1,1]}(4)\otimes S^0_{[1,1,-1,-1]}(4)\otimes S^0_{[1,-1,1,-1]}(4)\otimes S^0_{[1,-1,-1,1]}(4)\).
    Non-Gaussian uncertainties (open points) set in much earlier on this larger volume than the smaller volumes.
    However, central values and uncertainties sit lower all the way out to the critical potential.
    This is a significant improvement at large $\mu$ compared to the smaller volumes and to conventional Monte-Carlo resampling.
    Points have been offset slightly in $\mu$ for readability.}
    \label{fig:L16161616}
\end{figure}

On the largest (\(16\times16\times16\times16\)) lattice, the best approach we identified was again a combination of four orthogonal diagonal slicing patterns, with twice as many slices as on the smaller lattice: \(S^0_{[1,1,1,1]}(4)\otimes S^0_{[1,1,-1,-1]}(4)\otimes S^0_{[1,-1,1,-1]}(4)\otimes S^0_{[1,-1,-1,1]}(4)\), with \(N^{(0)} = 1,000\) and \(N^{(1)} = 1,000\).
Fig.~\ref{fig:L16161616} shows that this approach has the best-behaved values of all the volumes considered so far, all the way out to the phase transition.
However, the uncertainties are not as well-behaved, with the overlap problem setting in at a much smaller \(\mu=0.15\).
This isn't unique to the multi-level calculation, reweighting with conventional Monte-Carlo performs extremely poorly at this volume.

\section{Conclusion}
With an appropriate decomposition of the lattice into disjoint domains, a multi-level Monte-Carlo approach
allows practical simulations using reweighting at significantly higher chemical potentials than reweighting alone.
This is achieved through a numerically efficient cross-combination of independent Monte-Carlo ensembles,
providing a huge statistical advantage, as described in Section~\ref{sec:multi-level:formulation}.
The best approach we identified for the larger volumes we considered was a combination of orthogonal slicings \(S^0_{[1,1,1,1]}(D)\otimes S^0_{[1,1,-1,-1]}(D)\otimes S^0_{[1,-1,1,-1]}(D)\otimes S^0_{[1,-1,-1,1]}(D)\).
This approach provided a significant extension to the range of chemical potentials for which a precise result is attainable, and improved results all the way out to the phase transition,
for the same or better compute cost.
Our approach achieved results consistent with previous complex Langevin calculations~\cite{Aarts:2008wh,
Aarts:2009hn}.
This offers a promising approach to intermediate-to-high chemical potential calculations, that is free from complex
Langevin's unresolved convergence issues for general simulations.

However, the application of this approach to quantum chromodynamics (QCD) is complicated by long-range effects arising
from pions.
The frozen regions between blocks need to be much thicker than the single lattice sites used here.
Future work will investigate these decomposition approaches with thicker frozen regions between the
domains, capable of suppressing the problematic pion contributions.
It will be important to explore how these frozen regions, and the resulting performance of the
calculations needs to scale with pion mass, to identify if there is a way to controllably scale the
calculation to light pion masses with currently-available supercomputing resources.

Further, at higher chemical potentials, this approach ran into issues with overlap between the sampled phase-quenched distribution and the true underlying distribution.
This led to non-Gaussian behaviour in the statistical noise, making uncertainties unreliable.
Sign reweighting~\cite{Hsu:2010zza,deForcrand:2002pa,Borsanyi:2021hbk} is a proposed solution to this problem that eliminates the overlap problem and only leaves the sign problem, that this approach seems to resolve.
It will be valuable to explore how these two techniques can be combined to produce an approach free from both problems.
It will also be valuable to explore how the situation changes if multi-level sampling with more than two levels is used, as described at the end of Sec.~\ref{sec:multi-level:formulation}.

In conclusion, this is a promising result, and a sign that this approach may offer a pathway
to simulating QCD at nuclear matter densities or beyond, providing key theoretical insight into nuclear
and neutron star physics.

\ack
We warmly thank Kalman Szabo for valuable discussions and insight, as well as providing an example code
for multi-level implementation.
This research was supported with supercomputing resources provided by the
Phoenix HPC service at the University of Adelaide.
This research was undertaken with the assistance
of resources from the National Computational Infrastructure (NCI), provided through the National
Computational Merit Allocation Scheme
FS is supported by a Ramsay Fellowship from the University of Adelaide.
	
\section*{References}
\bibliographystyle{unsrturl}
\bibliography{multilevel}

@article{Aarts:2009hn,
    author = "Aarts, Gert",
    title = "{Complex Langevin dynamics at finite chemical potential: Mean field analysis in the relativistic Bose gas}",
    eprint = "0902.4686",
    archivePrefix = "arXiv",
    primaryClass = "hep-lat",
    doi = "10.1088/1126-6708/2009/05/052",
    journal = "JHEP",
    volume = "05",
    pages = "052",
    year = "2009"
}

@article{Aarts:2008wh,
    author = "Aarts, Gert",
    title = "{Can stochastic quantization evade the sign problem? The relativistic Bose gas at finite chemical potential}",
    eprint = "0810.2089",
    archivePrefix = "arXiv",
    primaryClass = "hep-lat",
    doi = "10.1103/PhysRevLett.102.131601",
    journal = "Phys. Rev. Lett.",
    volume = "102",
    pages = "131601",
    year = "2009"
}

@article{Luscher:2001up,
    author = "Luscher, Martin and Weisz, Peter",
    title = "{Locality and exponential error reduction in numerical lattice gauge theory}",
    eprint = "hep-lat/0108014",
    archivePrefix = "arXiv",
    reportNumber = "CERN-TH-2001-208, UT-CCP-P-110, MPI-PHT-2001-25",
    doi = "10.1088/1126-6708/2001/09/010",
    journal = "JHEP",
    volume = "09",
    pages = "010",
    year = "2001"
}

@article{Meyer:2002cd,
    author = "Meyer, Harvey B.",
    title = "{Locality and statistical error reduction on correlation functions}",
    eprint = "hep-lat/0209145",
    archivePrefix = "arXiv",
    doi = "10.1088/1126-6708/2003/01/048",
    journal = "JHEP",
    volume = "01",
    pages = "048",
    year = "2003"
}

@article{DellaMorte:2007zz,
    author = "Della Morte, Michele and Giusti, Leonardo",
    title = "{Exploiting symmetries for exponential error reduction in path integral Monte Carlo}",
    reportNumber = "CERN-PH-TH-2007-196",
    doi = "10.1016/j.cpc.2008.10.017",
    journal = "Comput. Phys. Commun.",
    volume = "180",
    pages = "813--818",
    year = "2009"
}

@article{DellaMorte:2008jd,
    author = "Della Morte, Michele and Giusti, Leonardo",
    title = "{Symmetries and exponential error reduction in Yang-Mills theories on the lattice}",
    eprint = "0806.2601",
    archivePrefix = "arXiv",
    primaryClass = "hep-lat",
    reportNumber = "CERN-PH-TH-2008-112",
    doi = "10.1016/j.cpc.2009.03.009",
    journal = "Comput. Phys. Commun.",
    volume = "180",
    pages = "819--826",
    year = "2009"
}

@article{DellaMorte:2010yp,
    author = "Della Morte, Michele and Giusti, Leonardo",
    title = "{A novel approach for computing glueball masses and matrix elements in Yang-Mills theories on the lattice}",
    eprint = "1012.2562",
    archivePrefix = "arXiv",
    primaryClass = "hep-lat",
    reportNumber = "CERN-PH-TH-2010-197, HIM-2010-02, MKPH-T-10-40",
    doi = "10.1007/JHEP05(2011)056",
    journal = "JHEP",
    volume = "05",
    pages = "056",
    year = "2011"
}

@article{Ce:2016idq,
    author = "C{\`e}, Marco and Giusti, Leonardo and Schaefer, Stefan",
    title = "{Domain decomposition, multi-level integration and exponential noise reduction in lattice QCD}",
    eprint = "1601.04587",
    archivePrefix = "arXiv",
    primaryClass = "hep-lat",
    reportNumber = "DESY-16-007",
    doi = "10.1103/PhysRevD.93.094507",
    journal = "Phys. Rev. D",
    volume = "93",
    number = "9",
    pages = "094507",
    year = "2016"
}

@article{Ce:2016qto,
    author = "C{\`e}, Marco and Giusti, Leonardo and Schaefer, Stefan",
    title = "{Domain decomposition and multilevel integration for fermions}",
    eprint = "1612.06424",
    archivePrefix = "arXiv",
    primaryClass = "hep-lat",
    reportNumber = "DESY-16-246",
    doi = "10.22323/1.256.0263",
    journal = "PoS",
    volume = "LATTICE2016",
    pages = "263",
    year = "2016"
}

@article{Troyer:2004ge,
    author = "Troyer, Matthias and Wiese, Uwe-Jens",
    title = "{Computational complexity and fundamental limitations to fermionic quantum Monte Carlo simulations}",
    eprint = "cond-mat/0408370",
    archivePrefix = "arXiv",
    doi = "10.1103/PhysRevLett.94.170201",
    journal = "Phys. Rev. Lett.",
    volume = "94",
    pages = "170201",
    year = "2005"
}

@article{Cohen:2003kd,
    author = "Cohen, Thomas D .",
    title = "{Functional integrals for QCD at nonzero chemical potential and zero density}",
    eprint = "hep-ph/0307089",
    archivePrefix = "arXiv",
    reportNumber = "DOE-ER-40762-287, UMDPP-03-058",
    doi = "10.1103/PhysRevLett.91.222001",
    journal = "Phys. Rev. Lett.",
    volume = "91",
    pages = "222001",
    year = "2003"
}

@article{Cristoforetti:2012su,
    author = "Cristoforetti, Marco and Di Renzo, Francesco and Scorzato, Luigi",
    collaboration = "AuroraScience",
    title = "{New approach to the sign problem in quantum field theories: High density QCD on a Lefschetz thimble}",
    eprint = "1205.3996",
    archivePrefix = "arXiv",
    primaryClass = "hep-lat",
    doi = "10.1103/PhysRevD.86.074506",
    journal = "Phys. Rev. D",
    volume = "86",
    pages = "074506",
    year = "2012"
}

@article{Cristoforetti:2013wha,
    author = "Cristoforetti, Marco and Di Renzo, Francesco and Mukherjee, Abhishek and Scorzato, Luigi",
    title = "{Monte Carlo simulations on the Lefschetz thimble: Taming the sign problem}",
    eprint = "1303.7204",
    archivePrefix = "arXiv",
    primaryClass = "hep-lat",
    doi = "10.1103/PhysRevD.88.051501",
    journal = "Phys. Rev. D",
    volume = "88",
    number = "5",
    pages = "051501",
    year = "2013"
}

@article{Alexandru:2015xva,
    author = {Alexandru, Andrei and Basar, G{\"o}k{\c{c}}e and Bedaque, Paulo},
    title = "{Monte Carlo algorithm for simulating fermions on Lefschetz thimbles}",
    eprint = "1510.03258",
    archivePrefix = "arXiv",
    primaryClass = "hep-lat",
    doi = "10.1103/PhysRevD.93.014504",
    journal = "Phys. Rev. D",
    volume = "93",
    number = "1",
    pages = "014504",
    year = "2016"
}

@article{Alexandru:2016lsn,
    author = "Alexandru, Andrei and Basar, Gokce and Bedaque, Paulo F. and Ridgway, Gregory W. and Warrington, Neill C.",
    title = "{Fast estimator of Jacobians in the Monte Carlo integration on Lefschetz thimbles}",
    eprint = "1604.00956",
    archivePrefix = "arXiv",
    primaryClass = "hep-lat",
    doi = "10.1103/PhysRevD.93.094514",
    journal = "Phys. Rev. D",
    volume = "93",
    number = "9",
    pages = "094514",
    year = "2016"
}

@article{Nishimura:2017vav,
    author = "Nishimura, Jun and Shimasaki, Shinji",
    title = "{Combining the complex Langevin method and the generalized Lefschetz-thimble method}",
    eprint = "1703.09409",
    archivePrefix = "arXiv",
    primaryClass = "hep-lat",
    reportNumber = "KEK-TH-1967",
    doi = "10.1007/JHEP06(2017)023",
    journal = "JHEP",
    volume = "06",
    pages = "023",
    year = "2017"
}

@article{Seiler:2012wz,
    author = "Seiler, Erhard and Sexty, Denes and Stamatescu, Ion-Olimpiu",
    title = "{Gauge cooling in complex Langevin for QCD with heavy quarks}",
    eprint = "1211.3709",
    archivePrefix = "arXiv",
    primaryClass = "hep-lat",
    reportNumber = "MPP-2012-150",
    doi = "10.1016/j.physletb.2013.04.062",
    journal = "Phys. Lett. B",
    volume = "723",
    pages = "213--216",
    year = "2013"
}

@article{Sexty:2013ica,
    author = "Sexty, D{\'e}nes",
    title = "{Simulating full QCD at nonzero density using the complex Langevin equation}",
    eprint = "1307.7748",
    archivePrefix = "arXiv",
    primaryClass = "hep-lat",
    doi = "10.1016/j.physletb.2014.01.019",
    journal = "Phys. Lett. B",
    volume = "729",
    pages = "108--111",
    year = "2014"
}

@article{Aarts:2014bwa,
    author = "Aarts, Gert and Seiler, Erhard and Sexty, D{\'e}nes and Stamatescu, Ion-Olimpiu",
    title = "{Simulating QCD at nonzero baryon density to all orders in the hopping parameter expansion}",
    eprint = "1408.3770",
    archivePrefix = "arXiv",
    primaryClass = "hep-lat",
    doi = "10.1103/PhysRevD.90.114505",
    journal = "Phys. Rev. D",
    volume = "90",
    number = "11",
    pages = "114505",
    year = "2014"
}

@article{Fodor:2015doa,
    author = {Fodor, Z. and Katz, S. D. and Sexty, D. and T{\"o}r{\"o}k, C.},
    title = "{Complex Langevin dynamics for dynamical QCD at nonzero chemical potential: A comparison with multiparameter reweighting}",
    eprint = "1508.05260",
    archivePrefix = "arXiv",
    primaryClass = "hep-lat",
    doi = "10.1103/PhysRevD.92.094516",
    journal = "Phys. Rev. D",
    volume = "92",
    number = "9",
    pages = "094516",
    year = "2015"
}

@article{Sexty:2019vqx,
    author = "Sexty, D{\'e}nes",
    title = "{Calculating the equation of state of dense quark-gluon plasma using the complex Langevin equation}",
    eprint = "1907.08712",
    archivePrefix = "arXiv",
    primaryClass = "hep-lat",
    doi = "10.1103/PhysRevD.100.074503",
    journal = "Phys. Rev. D",
    volume = "100",
    number = "7",
    pages = "074503",
    year = "2019"
}

@article{Kogut:2019qmi,
    author = "Kogut, J. B. and Sinclair, D. K.",
    title = "{Applying Complex Langevin Simulations to Lattice QCD at Finite Density}",
    eprint = "1903.02622",
    archivePrefix = "arXiv",
    primaryClass = "hep-lat",
    doi = "10.1103/PhysRevD.100.054512",
    journal = "Phys. Rev. D",
    volume = "100",
    number = "5",
    pages = "054512",
    year = "2019"
}

@article{Scherzer:2020kiu,
    author = "Scherzer, M. and Sexty, D. and Stamatescu, I. O.",
    title = "{Deconfinement transition line with the complex Langevin equation up to $\mu /T \sim 5$}",
    eprint = "2004.05372",
    archivePrefix = "arXiv",
    primaryClass = "hep-lat",
    doi = "10.1103/PhysRevD.102.014515",
    journal = "Phys. Rev. D",
    volume = "102",
    number = "1",
    pages = "014515",
    year = "2020"
}

@article{Barbour:1997ej,
    author = "Barbour, Ian M. and Morrison, Susan E. and Klepfish, Elyakum G. and Kogut, John B. and Lombardo, Maria-Paola",
    editor = "Iwasaki, Y. and Ukawa, A.",
    title = "{Results on finite density QCD}",
    eprint = "hep-lat/9705042",
    archivePrefix = "arXiv",
    doi = "10.1016/S0920-5632(97)00484-2",
    journal = "Nucl. Phys. B Proc. Suppl.",
    volume = "60",
    pages = "220--234",
    year = "1998"
}

@article{Fodor:2001au,
    author = "Fodor, Z. and Katz, S. D.",
    title = "{A New method to study lattice QCD at finite temperature and chemical potential}",
    eprint = "hep-lat/0104001",
    archivePrefix = "arXiv",
    reportNumber = "DESY-01-044, ITP-BUDAPEST-566",
    doi = "10.1016/S0370-2693(02)01583-6",
    journal = "Phys. Lett. B",
    volume = "534",
    pages = "87--92",
    year = "2002"
}

@article{Fodor:2001pe,
    author = "Fodor, Z. and Katz, S. D.",
    title = "{Lattice determination of the critical point of QCD at finite T and mu}",
    eprint = "hep-lat/0106002",
    archivePrefix = "arXiv",
    reportNumber = "ITP-BUDAPEST-568, DESY-01-057",
    doi = "10.1088/1126-6708/2002/03/014",
    journal = "JHEP",
    volume = "03",
    pages = "014",
    year = "2002"
}

@article{Fodor:2004nz,
    author = "Fodor, Z. and Katz, S. D.",
    title = "{Critical point of QCD at finite T and mu, lattice results for physical quark masses}",
    eprint = "hep-lat/0402006",
    archivePrefix = "arXiv",
    reportNumber = "ITP-BUDAPEST-609, WUB-04-04",
    doi = "10.1088/1126-6708/2004/04/050",
    journal = "JHEP",
    volume = "04",
    pages = "050",
    year = "2004"
}

@article{Giordano:2019gev,
    author = "Giordano, Matteo and Kapas, Kornel and Katz, Sandor D. and Nogradi, Daniel and Pasztor, Attila",
    title = "{Radius of convergence in lattice QCD at finite $\mu_B$ with rooted staggered fermions}",
    eprint = "1911.00043",
    archivePrefix = "arXiv",
    primaryClass = "hep-lat",
    doi = "10.1103/PhysRevD.101.074511",
    journal = "Phys. Rev. D",
    volume = "101",
    number = "7",
    pages = "074511",
    year = "2020",
    note = "[Erratum: Phys.Rev.D 104, 119901 (2021)]"
}

@article{deForcrand:2002hgr,
    author = "de Forcrand, Philippe and Philipsen, Owe",
    title = "{The QCD phase diagram for small densities from imaginary chemical potential}",
    eprint = "hep-lat/0205016",
    archivePrefix = "arXiv",
    reportNumber = "MIT-CTP-3270, CERN-TH-2002-102",
    doi = "10.1016/S0550-3213(02)00626-0",
    journal = "Nucl. Phys. B",
    volume = "642",
    pages = "290--306",
    year = "2002"
}

@article{DElia:2002tig,
    author = "D'Elia, Massimo and Lombardo, Maria-Paola",
    title = "{Finite density QCD via imaginary chemical potential}",
    eprint = "hep-lat/0209146",
    archivePrefix = "arXiv",
    reportNumber = "GEF-TH-2002-12",
    doi = "10.1103/PhysRevD.67.014505",
    journal = "Phys. Rev. D",
    volume = "67",
    pages = "014505",
    year = "2003"
}

@article{DElia:2009pdy,
    author = "D'Elia, Massimo and Sanfilippo, Francesco",
    title = "{Thermodynamics of two flavor QCD from imaginary chemical potentials}",
    eprint = "0904.1400",
    archivePrefix = "arXiv",
    primaryClass = "hep-lat",
    reportNumber = "GEF-TH-01-09",
    doi = "10.1103/PhysRevD.80.014502",
    journal = "Phys. Rev. D",
    volume = "80",
    pages = "014502",
    year = "2009"
}

@article{Cea:2014xva,
    author = "Cea, Paolo and Cosmai, Leonardo and Papa, Alessandro",
    title = "{Critical line of 2+1 flavor QCD}",
    eprint = "1403.0821",
    archivePrefix = "arXiv",
    primaryClass = "hep-lat",
    doi = "10.1103/PhysRevD.89.074512",
    journal = "Phys. Rev. D",
    volume = "89",
    number = "7",
    pages = "074512",
    year = "2014"
}

@article{Bonati:2014kpa,
    author = "Bonati, Claudio and de Forcrand, Philippe and D'Elia, Massimo and Philipsen, Owe and Sanfilippo, Francesco",
    title = "{Chiral phase transition in two-flavor QCD from an imaginary chemical potential}",
    eprint = "1408.5086",
    archivePrefix = "arXiv",
    primaryClass = "hep-lat",
    reportNumber = "CERN-PH-TH-2014-158, IFUP-TH-2014-8",
    doi = "10.1103/PhysRevD.90.074030",
    journal = "Phys. Rev. D",
    volume = "90",
    number = "7",
    pages = "074030",
    year = "2014"
}

@article{Cea:2015cya,
    author = "Cea, Paolo and Cosmai, Leonardo and Papa, Alessandro",
    title = "{Critical line of 2+1 flavor QCD: Toward the continuum limit}",
    eprint = "1508.07599",
    archivePrefix = "arXiv",
    primaryClass = "hep-lat",
    doi = "10.1103/PhysRevD.93.014507",
    journal = "Phys. Rev. D",
    volume = "93",
    number = "1",
    pages = "014507",
    year = "2016"
}

@article{Bonati:2015bha,
    author = "Bonati, Claudio and D'Elia, Massimo and Mariti, Marco and Mesiti, Michele and Negro, Francesco and Sanfilippo, Francesco",
    title = "{Curvature of the chiral pseudocritical line in QCD: Continuum extrapolated results}",
    eprint = "1507.03571",
    archivePrefix = "arXiv",
    primaryClass = "hep-lat",
    reportNumber = "IFUP-TH-2015-07",
    doi = "10.1103/PhysRevD.92.054503",
    journal = "Phys. Rev. D",
    volume = "92",
    number = "5",
    pages = "054503",
    year = "2015"
}

@article{Bellwied:2015rza,
    author = {Bellwied, R. and Borsanyi, S. and Fodor, Z. and G{\"u}nther, J. and Katz, S. D. and Ratti, C. and Szabo, K. K.},
    title = "{The QCD phase diagram from analytic continuation}",
    eprint = "1507.07510",
    archivePrefix = "arXiv",
    primaryClass = "hep-lat",
    doi = "10.1016/j.physletb.2015.11.011",
    journal = "Phys. Lett. B",
    volume = "751",
    pages = "559--564",
    year = "2015"
}

@article{DElia:2016jqh,
    author = "D'Elia, Massimo and Gagliardi, Giuseppe and Sanfilippo, Francesco",
    title = "{Higher order quark number fluctuations via imaginary chemical potentials in $N_f=2+1$ QCD}",
    eprint = "1611.08285",
    archivePrefix = "arXiv",
    primaryClass = "hep-lat",
    doi = "10.1103/PhysRevD.95.094503",
    journal = "Phys. Rev. D",
    volume = "95",
    number = "9",
    pages = "094503",
    year = "2017"
}

@article{Guenther:2017hnx,
    author = "Guenther, J. N. and Bellwied, R. and Borsanyi, S. and Fodor, Z. and Katz, S. D. and Pasztor, A. and Ratti, C. and Szab{\'o}, K. K.",
    editor = "Heinz, Ulrich and Evdokimov, Olga and Jacobs, Peter",
    title = "{The QCD equation of state at finite density from analytical continuation}",
    eprint = "1607.02493",
    archivePrefix = "arXiv",
    primaryClass = "hep-lat",
    doi = "10.1016/j.nuclphysa.2017.05.044",
    journal = "Nucl. Phys. A",
    volume = "967",
    pages = "720--723",
    year = "2017"
}

@article{Alba:2017mqu,
    author = "Alba, Paolo and others",
    title = "{Constraining the hadronic spectrum through QCD thermodynamics on the lattice}",
    eprint = "1702.01113",
    archivePrefix = "arXiv",
    primaryClass = "hep-lat",
    doi = "10.1103/PhysRevD.96.034517",
    journal = "Phys. Rev. D",
    volume = "96",
    number = "3",
    pages = "034517",
    year = "2017"
}

@article{Vovchenko:2017xad,
    author = "Vovchenko, Volodymyr and Pasztor, Attila and Fodor, Zoltan and Katz, Sandor D. and Stoecker, Horst",
    title = "{Repulsive baryonic interactions and lattice QCD observables at imaginary chemical potential}",
    eprint = "1708.02852",
    archivePrefix = "arXiv",
    primaryClass = "hep-ph",
    doi = "10.1016/j.physletb.2017.10.042",
    journal = "Phys. Lett. B",
    volume = "775",
    pages = "71--78",
    year = "2017"
}

@article{Bonati:2018nut,
    author = "Bonati, Claudio and D'Elia, Massimo and Negro, Francesco and Sanfilippo, Francesco and Zambello, Kevin",
    title = "{Curvature of the pseudocritical line in QCD: Taylor expansion matches analytic continuation}",
    eprint = "1805.02960",
    archivePrefix = "arXiv",
    primaryClass = "hep-lat",
    doi = "10.1103/PhysRevD.98.054510",
    journal = "Phys. Rev. D",
    volume = "98",
    number = "5",
    pages = "054510",
    year = "2018"
}

@article{Borsanyi:2018grb,
    author = "Borsanyi, Szabolcs and Fodor, Zoltan and Guenther, Jana N. and Katz, Sandor K. and Szabo, Kalman K. and Pasztor, Attila and Portillo, Israel and Ratti, Claudia",
    title = "{Higher order fluctuations and correlations of conserved charges from lattice QCD}",
    eprint = "1805.04445",
    archivePrefix = "arXiv",
    primaryClass = "hep-lat",
    doi = "10.1007/JHEP10(2018)205",
    journal = "JHEP",
    volume = "10",
    pages = "205",
    year = "2018"
}

@article{Bellwied:2019pxh,
    author = "Bellwied, Rene and Borsanyi, Szabolcs and Fodor, Zoltan and Guenther, Jana N. and Noronha-Hostler, Jacquelyn and Parotto, Paolo and Pasztor, Attila and Ratti, Claudia and Stafford, Jamie M.",
    title = "{Off-diagonal correlators of conserved charges from lattice QCD and how to relate them to experiment}",
    eprint = "1910.14592",
    archivePrefix = "arXiv",
    primaryClass = "hep-lat",
    doi = "10.1103/PhysRevD.101.034506",
    journal = "Phys. Rev. D",
    volume = "101",
    number = "3",
    pages = "034506",
    year = "2020"
}

@article{Borsanyi:2020fev,
    author = "Borsanyi, Szabolcs and Fodor, Zoltan and Guenther, Jana N. and Kara, Ruben and Katz, Sandor D. and Parotto, Paolo and Pasztor, Attila and Ratti, Claudia and Szabo, Kalman K.",
    title = "{QCD Crossover at Finite Chemical Potential from Lattice Simulations}",
    eprint = "2002.02821",
    archivePrefix = "arXiv",
    primaryClass = "hep-lat",
    doi = "10.1103/PhysRevLett.125.052001",
    journal = "Phys. Rev. Lett.",
    volume = "125",
    number = "5",
    pages = "052001",
    year = "2020"
}

@article{Allton:2002zi,
    author = "Allton, C. R. and Ejiri, S. and Hands, S. J. and Kaczmarek, O. and Karsch, F. and Laermann, E. and Schmidt, C. and Scorzato, L.",
    title = "{The QCD thermal phase transition in the presence of a small chemical potential}",
    eprint = "hep-lat/0204010",
    archivePrefix = "arXiv",
    reportNumber = "SWAT-02-335, NSF-ITP-02-26, BI-TP-2002-06",
    doi = "10.1103/PhysRevD.66.074507",
    journal = "Phys. Rev. D",
    volume = "66",
    pages = "074507",
    year = "2002"
}

@article{Gavai:2003mf,
    author = "Gavai, Rajiv V. and Gupta, Sourendu",
    title = "{Pressure and nonlinear susceptibilities in QCD at finite chemical potentials}",
    eprint = "hep-lat/0303013",
    archivePrefix = "arXiv",
    reportNumber = "TIFR-TH-03-07",
    doi = "10.1103/PhysRevD.68.034506",
    journal = "Phys. Rev. D",
    volume = "68",
    pages = "034506",
    year = "2003"
}

@article{Gavai:2004sd,
    author = "Gavai, R. V. and Gupta, Sourendu",
    title = "{The Critical end point of QCD}",
    eprint = "hep-lat/0412035",
    archivePrefix = "arXiv",
    doi = "10.1103/PhysRevD.71.114014",
    journal = "Phys. Rev. D",
    volume = "71",
    pages = "114014",
    year = "2005"
}

@article{Allton:2005gk,
    author = "Allton, C. R. and Doring, M. and Ejiri, S. and Hands, S. J. and Kaczmarek, O. and Karsch, F. and Laermann, E. and Redlich, K.",
    title = "{Thermodynamics of two flavor QCD to sixth order in quark chemical potential}",
    eprint = "hep-lat/0501030",
    archivePrefix = "arXiv",
    reportNumber = "BI-TP-2005-02, SWAT-05-422",
    doi = "10.1103/PhysRevD.71.054508",
    journal = "Phys. Rev. D",
    volume = "71",
    pages = "054508",
    year = "2005"
}

@article{Gavai:2008zr,
    author = "Gavai, R. V. and Gupta, Sourendu",
    title = "{QCD at finite chemical potential with six time slices}",
    eprint = "0806.2233",
    archivePrefix = "arXiv",
    primaryClass = "hep-lat",
    reportNumber = "TIFR-TH-08-26",
    doi = "10.1103/PhysRevD.78.114503",
    journal = "Phys. Rev. D",
    volume = "78",
    pages = "114503",
    year = "2008"
}

@article{MILC:2008reg,
    author = "Basak, S. and others",
    editor = "Aubin, Christopher and Cohen, Saul and Dawson, Chris and Dudek, Jozef and Edwards, Robert and Joo, Balint and Lin, Huey-Wen and Orginos, Kostas and Richards, David and Thacker, Hank",
    collaboration = "MILC",
    title = "{QCD equation of state at non-zero chemical potential}",
    eprint = "0910.0276",
    archivePrefix = "arXiv",
    primaryClass = "hep-lat",
    doi = "10.22323/1.066.0171",
    journal = "PoS",
    volume = "LATTICE2008",
    pages = "171",
    year = "2008"
}

@article{Borsanyi:2011sw,
    author = "Borsanyi, Szabolcs and Fodor, Zoltan and Katz, Sandor D. and Krieg, Stefan and Ratti, Claudia and Szabo, Kalman",
    title = "{Fluctuations of conserved charges at finite temperature from lattice QCD}",
    eprint = "1112.4416",
    archivePrefix = "arXiv",
    primaryClass = "hep-lat",
    doi = "10.1007/JHEP01(2012)138",
    journal = "JHEP",
    volume = "01",
    pages = "138",
    year = "2012"
}

@article{Borsanyi:2012cr,
    author = "Borsanyi, Sz. and Endrodi, G. and Fodor, Z. and Katz, S. D. and Krieg, S. and Ratti, C. and Szabo, K. K.",
    title = "{QCD equation of state at nonzero chemical potential: continuum results with physical quark masses at order $mu^2$}",
    eprint = "1204.6710",
    archivePrefix = "arXiv",
    primaryClass = "hep-lat",
    reportNumber = "WUB-12-10",
    doi = "10.1007/JHEP08(2012)053",
    journal = "JHEP",
    volume = "08",
    pages = "053",
    year = "2012"
}

@article{Bellwied:2015lba,
    author = "Bellwied, R. and Borsanyi, S. and Fodor, Z. and Katz, S. D. and Pasztor, A. and Ratti, C. and Szabo, K. K.",
    title = "{Fluctuations and correlations in high temperature QCD}",
    eprint = "1507.04627",
    archivePrefix = "arXiv",
    primaryClass = "hep-lat",
    doi = "10.1103/PhysRevD.92.114505",
    journal = "Phys. Rev. D",
    volume = "92",
    number = "11",
    pages = "114505",
    year = "2015"
}

@article{Ding:2015fca,
    author = "Ding, H. -T. and Mukherjee, Swagato and Ohno, H. and Petreczky, P. and Schadler, H. -P.",
    title = "{Diagonal and off-diagonal quark number susceptibilities at high temperatures}",
    eprint = "1507.06637",
    archivePrefix = "arXiv",
    primaryClass = "hep-lat",
    doi = "10.1103/PhysRevD.92.074043",
    journal = "Phys. Rev. D",
    volume = "92",
    number = "7",
    pages = "074043",
    year = "2015"
}

@article{Bazavov:2017dus,
    author = "Bazavov, A. and others",
    title = "{The QCD Equation of State to $\mathcal{O}(\mu_B^6)$ from Lattice QCD}",
    eprint = "1701.04325",
    archivePrefix = "arXiv",
    primaryClass = "hep-lat",
    doi = "10.1103/PhysRevD.95.054504",
    journal = "Phys. Rev. D",
    volume = "95",
    number = "5",
    pages = "054504",
    year = "2017"
}

@article{HotQCD:2018pds,
    author = "Bazavov, A. and others",
    collaboration = "HotQCD",
    title = "{Chiral crossover in QCD at zero and non-zero chemical potentials}",
    eprint = "1812.08235",
    archivePrefix = "arXiv",
    primaryClass = "hep-lat",
    doi = "10.1016/j.physletb.2019.05.013",
    journal = "Phys. Lett. B",
    volume = "795",
    pages = "15--21",
    year = "2019"
}

@article{Giordano:2019slo,
    author = "Giordano, Matteo and P{\'a}sztor, Attila",
    title = "{Reliable estimation of the radius of convergence in finite density QCD}",
    eprint = "1904.01974",
    archivePrefix = "arXiv",
    primaryClass = "hep-lat",
    doi = "10.1103/PhysRevD.99.114510",
    journal = "Phys. Rev. D",
    volume = "99",
    number = "11",
    pages = "114510",
    year = "2019"
}

@article{Bazavov:2020bjn,
    author = "Bazavov, A. and others",
    title = "{Skewness, kurtosis, and the fifth and sixth order cumulants of net baryon-number distributions from lattice QCD confront high-statistics STAR data}",
    eprint = "2001.08530",
    archivePrefix = "arXiv",
    primaryClass = "hep-lat",
    doi = "10.1103/PhysRevD.101.074502",
    journal = "Phys. Rev. D",
    volume = "101",
    number = "7",
    pages = "074502",
    year = "2020"
}

@article{Tsutsui:2025jez,
    author = "Tsutsui, Shoichiro and Asano, Yuhma and Ito, Yuta and Matsufuru, Hideo and Namekawa, Yusuke and Nishimura, Jun and Shimasaki, Shinji and Tsuchiya, Asato",
    title = "{On the validity of the complex Langevin method near the deconfining phase transition in QCD at finite density}",
    eprint = "2505.06551",
    archivePrefix = "arXiv",
    primaryClass = "hep-lat",
    reportNumber = "KEK-TH-2355, UTHEP-801, RIKEN-iTHEMS-Report-25",
    doi = "10.1007/JHEP10(2025)108",
    journal = "JHEP",
    volume = "10",
    pages = "108",
    year = "2025"
}

@article{Ito:2020mys,
    author = "Ito, Yuta and Matsufuru, Hideo and Namekawa, Yusuke and Nishimura, Jun and Shimasaki, Shinji and Tsuchiya, Asato and Tsutsui, Shoichiro",
    title = "{Complex Langevin calculations in QCD at finite density}",
    eprint = "2007.08778",
    archivePrefix = "arXiv",
    primaryClass = "hep-lat",
    reportNumber = "KEK-TH-2230, RIKEN-QHP-479",
    doi = "10.1007/JHEP10(2020)144",
    journal = "JHEP",
    volume = "10",
    pages = "144",
    year = "2020"
}

@article{Lombardo:2006yc,
    author = "Lombardo, M. P.",
    editor = "Becattini, Francesco",
    title = "{Lattice QCD at finite density: Imaginary chemical potential}",
    eprint = "hep-lat/0612017",
    archivePrefix = "arXiv",
    doi = "10.22323/1.029.0003",
    journal = "PoS",
    volume = "CPOD2006",
    pages = "003",
    year = "2006"
}

@article{Cohen:2025ahp,
    author = "Cohen, Thomas D.",
    title = "{Method to Obtain Bounds on the Equation of State of Cold Nuclear Matter from Imaginary Chemical Potentials}",
    eprint = "2510.07124",
    archivePrefix = "arXiv",
    primaryClass = "nucl-th",
    doi = "10.1103/7zxr-j397",
    journal = "Phys. Rev. Lett.",
    volume = "136",
    number = "18",
    pages = "181901",
    year = "2026"
}

@article{Borsanyi:2024xrx,
    author = "Borsanyi, Szabolcs and Fodor, Zoltan and Guenther, Jana N. and Parotto, Paolo and Pasztor, Attila and Pirelli, Ludovica and Szabo, Kalman K. and Wong, Chik Him",
    title = "{QCD deconfinement transition line up to {\ensuremath{\mu}}B=400{\,}{\,}MeV from finite volume lattice simulations}",
    eprint = "2410.06216",
    archivePrefix = "arXiv",
    primaryClass = "hep-lat",
    doi = "10.1103/PhysRevD.110.114507",
    journal = "Phys. Rev. D",
    volume = "110",
    number = "11",
    pages = "114507",
    year = "2024"
}

@article{Borsanyi:2021hbk,
    author = "Borsanyi, Szabolcs and Fodor, Zoltan and Giordano, Matteo and Katz, Sandor D. and Nogradi, Daniel and Pasztor, Attila and Wong, Chik Him",
    title = "{Lattice simulations of the QCD chiral transition at real baryon density}",
    eprint = "2108.09213",
    archivePrefix = "arXiv",
    primaryClass = "hep-lat",
    doi = "10.1103/PhysRevD.105.L051506",
    journal = "Phys. Rev. D",
    volume = "105",
    number = "5",
    pages = "L051506",
    year = "2022"
}

@article{Osborn:2004rf,
    author = "Osborn, James C.",
    title = "{Universal results from an alternate random matrix model for QCD with a baryon chemical potential}",
    eprint = "hep-th/0403131",
    archivePrefix = "arXiv",
    doi = "10.1103/PhysRevLett.93.222001",
    journal = "Phys. Rev. Lett.",
    volume = "93",
    pages = "222001",
    year = "2004"
}

@article{Osborn:2005ss,
    author = "Osborn, J. C. and Splittorff, K. and Verbaarschot, J. J. M.",
    title = "{Chiral symmetry breaking and the Dirac spectrum at nonzero chemical potential}",
    eprint = "hep-th/0501210",
    archivePrefix = "arXiv",
    reportNumber = "SUNY-NTG-05-1",
    doi = "10.1103/PhysRevLett.94.202001",
    journal = "Phys. Rev. Lett.",
    volume = "94",
    pages = "202001",
    year = "2005"
}

@article{Akemann:2004dr,
    author = "Akemann, G. and Osborn, J. C. and Splittorff, K. and Verbaarschot, J. J. M.",
    title = "{Unquenched QCD Dirac operator spectra at nonzero baryon chemical potential}",
    eprint = "hep-th/0411030",
    archivePrefix = "arXiv",
    reportNumber = "SUNY-NTG-04-03, NORDITA-2004-78-HE",
    doi = "10.1016/j.nuclphysb.2005.01.018",
    journal = "Nucl. Phys. B",
    volume = "712",
    pages = "287--324",
    year = "2005"
}

@article{Vladikas:1987tt,
    author = "Vladikas, A.",
    editor = "Billoire, A. and Lacaze, R. and Morel, A. and Napoly, O. and Zinn-Justin, Jean",
    title = "{QCD AT FINITE BARYON DENSITY}",
    reportNumber = "LNF-87/97-PT",
    doi = "10.1016/0920-5632(88)90122-3",
    journal = "Nucl. Phys. B Proc. Suppl.",
    volume = "4",
    pages = "322",
    year = "1988"
}

@article{Hsu:2010zza,
    author = "Hsu, Stephen D. H. and Reeb, David",
    title = "{On the sign problem in dense QCD}",
    doi = "10.1142/S0217751X10047968",
    journal = "Int. J. Mod. Phys. A",
    volume = "25",
    pages = "53--67",
    year = "2010"
}

@article{deForcrand:2002pa,
    author = "de Forcrand, P. and Kim, S. and Takaishi, T.",
    editor = "Edwards, R. and Negele, John W. and Richards, D.",
    title = "{QCD simulations at small chemical potential}",
    eprint = "hep-lat/0209126",
    archivePrefix = "arXiv",
    doi = "10.1016/S0920-5632(03)80451-6",
    journal = "Nucl. Phys. B Proc. Suppl.",
    volume = "119",
    pages = "541--543",
    year = "2003"
}

\end{document}